\documentclass[twocolumn, prl]{revtex4}
\usepackage{amsmath}
\usepackage{amssymb}
\usepackage{epsfig}

\newcommand{\ket}[1]{| #1 \rangle}
\newcommand{\bra}[1]{\langle #1 |}
\newcommand{\proj}[1]{\ket{#1}\!\bra{#1}}

\newcommand{\M}{\mathcal{M}}
\newcommand{\T}{\mathcal{T}}
\newcommand{\MT}{\tilde{\mathcal{M}}}
\newcommand{\np}{\left(1 \! - \! p \right)}
\newcommand{\ot}{\mathrm{otherwise}}

\begin{document}

\title{No purification for two copies of a noisy entangled state}

\author{Anthony J. Short}
\affiliation{DAMTP, Centre for Mathematical Sciences, Wilberforce
Road, Cambridge CB3 0WA, UK }

\begin{abstract}
We consider whether two copies of a noisy entangled state can be
transformed into a single copy of greater purity using local
operations and classical communication. We show that it is never
possible to achieve such a purification with certainty when the
family of noisy states is twirlable (i.e. when there exists a local
transformation that maps all states into the family, yet leaves the
family itself invariant). This implies that two copies of a Werner
state cannot be deterministically purified. Furthermore, due to the
construction of the proof, it will hold not only in quantum theory,
but in any generalised probabilistic theory. We use this to show that two
copies of a noisy PR-box (a hypothetical device more non-local than
is allowed by quantum theory) cannot be purified.

\end{abstract}

\maketitle

The ability to purify entanglement is a crucial feature of quantum
theory, allowing imperfect states to be refined into those necessary
to correctly implement quantum information protocols. Here we
consider whether two copies of a noisy entangled
state can be transformed into a single copy of greater purity, using local
operations and classical communication (LOCC). For the purposes of this paper,
we shall restrict our attention to purification protocols which work with certainty
(i.e. without post-selection).

We will show that two-copy purification is impossible to achieve
whenever the family of noisy entangled states is \emph{twirlable}
\cite{twirling}. That is, whenever there exists an LOCC
transformation that maps all states into the family, yet leaves the
family itself invariant. Twirling was first studied for Werner
states \cite{werner}, where it can be implemented by applying an
identical random unitary to both qubits. Our result therefore
implies that two copies of a Werner state cannot be
purified (although in \cite{werner} it is shown that two copies can
be purified using post-selection).

Aside from twirlability, the proof relies only on very weak
assumptions that are not specifically quantum. Indeed it will apply
in \emph{any} reasonable theory admitting mixed states and
entanglement, such as those defined in the general operational
framework of \cite{barrett}. It is interesting  that non-trivial results can be proved within such a general framework.
Furthermore, identifying which features are important in the proof allows a deeper and simpler understanding of it, even if one is only concerned with
quantum theory.

Recently however, much interest has focussed on PR-boxes \cite{pr, boxes}, hypothetical devices which are more non-local
than any quantum state, achieving the maximal possible violation of
the CHSH inequality \cite{chsh}. Many information theoretic
properties of PR-boxes have been studied. For example, it
has been shown that they would allow any communication complexity
problem to be solved with one bit of communication \cite{van-dam}.
Here, our proof implies that two noisy PR-boxes cannot be
deterministically purified.

 In our proof, we will use the same notation as the density matrix formalism of quantum theory. However, note that this notation is sufficient to describe a general  probabilistic theory. In particular, we will denote a mixture of states $s_i$ with probabilities $p_i$ by $s=\sum_i p_i s_i$,  a bipartite state in which the first part is in state $s_1$ and the second part is in state $s_2$ by $s_1 \otimes s_2$ (although we will not make use of the tensor product structure), and a transformation by $s' = \mathcal{T} [ s]$.

The proof only requires three reasonable assumptions of our physical theory, that are common to quantum theory, and the generalised probabilistic framework of \cite{barrett}.

\begin{enumerate}

\item \textbf{Existence of a separable state}: For the type of system considered, we assume that there exists at least one separable (i.e. un-entangled) state.   \label{one_sep_state}

\item  \textbf{Transformations act linearly on mixed states}: As probabilities may reflect a lack of knowledge about the state, rather than anything physical, we demand that transformations act linearly on mixed states. I.e. $\mathcal{T}[\sum_i p_i s_i] = \sum_i p_i \mathcal{T}[s_i]$.  \cite{barrett} \label{linearity}

\item \textbf{Entanglement cannot be created by LOCC}: We require that a separable state cannot be transformed into an entangled one via LOCC. This follows from any reasonable definition of local operations and entanglement. \label{entanglement}

\end{enumerate}

We now proceed to the proof of our main result, that two copies of a noisy entangled state from a twirlable family cannot be purified.

Consider a family of bipartite mixed states $s(p) \in S$ of the form:
\begin{equation}
s(p) = p s_0 + \np s_1 \quad p\in [0,1]
\end{equation}
where $s_0$ is a desired entangled state, and $s_1$ is the noise. As
discussed above, we will assume that this family of states is
twirlable.

\begin{enumerate}  \addtocounter{enumi}{3}

\item \textbf{Twirlability:} \label{twirlibility} A family of states $S$ is twirlable if there exists an LOCC transformation
$\mathcal{T}$ which leaves all states in $S$ invariant, and maps all allowed states into $S$.

\end{enumerate}
As twirling cannot transform a separable state into an entangled one (assumption \ref{entanglement}), and there exists at least
one separable initial state (assumption \ref{one_sep_state}), $S$ must also contain a separable state.
We denote the maximal value of $p$ for which $s(p)$ is separable by $p_s$ \footnote{For simplicity, we assume that the set of separable states is closed. If it is an open set, we take $p_s$ to be maximal boundary point of the set of $p$ for which $s(p)$ is separable, and the proof follows from the continuity of $Q(p)$.}.

We say that \emph{deterministic two-copy entanglement purification} of the family $S$ is
possible if there exists an LOCC transformation $\M$ such that
$\M[s(p) \otimes s(p) ] = s(p')$, for some $p,p' \in [0,1]$ satisfying $p' > p >
p_s$.

Given any transformation $\M$, we can always implement the twirled
transformation $\MT$ which consists of
carrying out $\M$ and then twirling (i.e. $\MT = \T \cdot \M$).
Furthermore, as twirling leaves all states in $S$ invariant, if $\M$
achieves a purification then so will $\MT$. Without loss of
generality, we therefore restrict our attention to twirled
transformations.

Consider applying a particular twirled transformation $\MT$ to the
states $s_i \otimes s_j$, $i,j \in [0,1]$. Due to twirlability (assumption \ref{twirlibility}), each
must be mapped to some state in $S$, and we denote the corresponding
purities by $q_{ij} \in [0,1]$:
\begin{equation}
\MT[s_i \otimes s_j] = s(q_{ij})
\end{equation}
From assumption \ref{linearity}, transformations must act linearly on probabilistic mixtures, hence
\begin{eqnarray}
s(p')\!\! &=& \! \MT\left[ \left(p s_0 + \np s_1 \right) \otimes  \left(p s_0 + \np s_1 \right)  \right] \nonumber \\
    &=& \! p^2 \MT[s_0 \otimes s_1] + p \np \MT[ s_0 \otimes s_1] \nonumber \\
        & & + p \np \MT[ s_1 \otimes s_0] + \np^2 \MT[s_1 \otimes s_1] \nonumber \\
    &=& \! p^2 s(q_{00}) + p\np s(q_{01}) \nonumber \\
        & & + p \np s(q_{10})  + \np^2 s(q_{11}) \nonumber \\
    &=& \! s(p^2 q_{00} + p \np (q_{01} + q_{10}) + \np^2 q_{11})
\end{eqnarray}

\begin{figure}
\epsfig{file=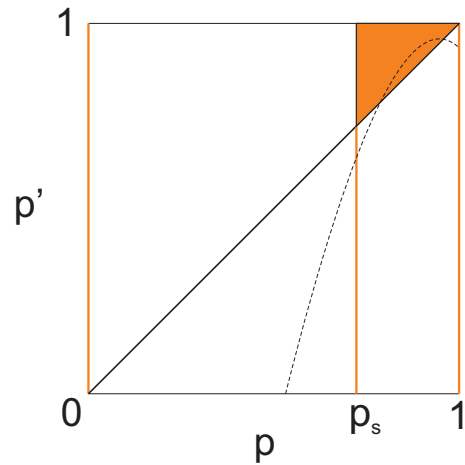, width=6cm} \caption{To achieve
purification, a function $p'=Q(p)$ must be quadratic, and pass
through all orange regions. This is impossible. The dashed line
shows one failed attempt, which is not universal (e.g. it does not
give $Q(0) \in [0,1]$) } \label{quadratic_fig}
\end{figure}

Writing
\begin{equation} \label{quadratic_eqn}
 p'=Q(p) =p^2 q_{00} + p \np (q_{01} + q_{10}) + \np^2 q_{11},
  \end{equation}
  we now establish four properties of the function $Q$, that are necessary to achieve a deterministic two-copy entanglement purification:

{   \renewcommand{\labelenumi}{\Alph{enumi}}
\begin{enumerate}

\item \textbf{Universal}: Because all input states are mapped into $S$ under $\MT$,  $ Q(p) \in [0,1]$ for all $p \in [0,1]$.

\item \textbf{Separability-preserving}: It must be the case that $Q(p_s) \leq p_s$, otherwise one would be able to transform the separable state $s(p_s)$ into an entangled state via LOCC (which would violate assumption \ref{entanglement}).

\item \textbf{Useful}. In order to achieve a useful purification of the state, there must exist some $p_e$ such that $Q(p_e) > p_e > p_s$.

\item \textbf{Quadratic}. From equation (\ref{quadratic_eqn}): $Q(p)$ is a quadratic (and hence continuous) function of $p$.

\end{enumerate} }
However, we now show that no function $Q(p)$ satisfying these four
requirements exists.
From requirements A-C above, we obtain four relations between $Q(p)$ and
$p$, for increasing values of $p$:
\begin{equation}
Q(0) \geq 0, \; Q(p_s) \leq p_s, \; Q(p_e) > p_e, \; \textrm{and} \;
Q(1) \leq 1
 \end{equation}
As $Q(p)$ is continuous, it is clear from these conditions that the
function $p'=Q(p)$ must intersect $p'=p$ at 3 or more points in the
interval $[0,1]$. The only
quadratic function to achieve this is $Q(p) = p$, but this is ruled
out by the third relation, $ Q(p_e) > p_e$. There are therefore no
functions $Q(p)$ obeying all  four necessary conditions (see figure \ref{quadratic_fig}).
Consequently, deterministic two-copy entanglement purification is
impossible for twirlable families of states.

\smallskip

\textbf{Werner states:} A important example of a twirlable family of entangled
states are the Werner states \cite{werner} in quantum theory. Here $s_0$ and $s_1$ are the two-qubit density matrices
\begin{eqnarray}
s_0 &=& \proj{\psi^-} = \frac{1}{2}\left(\ket{01}-\ket{10}\right)\left(\bra{01}-\bra{10}\right)\\
s_1 &=& \frac{1}{3} \left(\openone - \proj{\psi^-} \right)
\end{eqnarray}
The Werner states have $p_s=\frac{1}{2}$ (they are entangled for
$p>\frac{1}{2}$ and separable otherwise). The corresponding twirling
operation consists of applying the same randomly chosen unitary to
both qubits \cite{twirling}. i.e.
\begin{equation}
s \rightarrow \int (U \otimes U) s (U \otimes U)^{\dagger} dU
\end{equation}
where the integral is taken according to the (unitarily-invariant)
Haar measure \footnote{It is also possible to achieve this twirling by choosing a random unitary from a finite set corresponding to a
unitary 2-design  \cite{two-design}.}.

As they form a twirlable family of states, the above proof implies
that two copies of an entangled Werner state cannot be
deterministically purified.

Interestingly, it was shown by Bennett et al. \cite{twirling} that
two copies of an entangled Werner state \emph{can} be purified using
post-selection. This also implies that three (or more) copies of a
Werner state can be deterministically purified. One simply applies
the protocol given in  \cite{twirling} to the first two copies. If
the post-selection succeeds, the resultant higher-purity state is
output, and if the post-selection fails, the third copy is output
(with the original purity). This achieves
\begin{equation}
p' =  \frac{1}{9} \left( - 8 p^3 + 14p^2 +2p +1 \right),
\end{equation}
which satisfies $p' > p$ whenever $1>p>p_s$. Note that having three
copies of the state allows $p'$ to be a cubic function of  $p$,
which is able to satisfy  requirements A-C given above.

\smallskip

\textbf{Noisy PR-boxes:} The ability to generate `non-local'
correlations (i.e. correlations that cannot be explained by any
local hidden variable model \cite{chsh, bell}), is one of the most
surprising aspects of quantum theory. However, it is possible to
consider hypothetical systems that yield even stronger non-local
correlations than those attainable in quantum theory \cite{pr}, yet
which still cannot be used to signal.

The simplest devices of this type are known as PR-boxes. These are
composed of two terminals, each of which takes a binary input and
emits a binary output. Denoting the inputs by $x$ and $y$ and the
corresponding outputs by $a$ and $b$, the behavior of the PR-box is
characterised by the conditional probability distribution:
\begin{equation} \
P_{PR} (a b | x y) = \left\{
\begin{array}{ccl} \frac{1}{2} & : & a
\oplus  b = x y\; \textrm{(mod 2)} \\
0& :  & \ot
\end{array} \right. ,
\end{equation}
where $\oplus$ denotes addition modulo 2. With a PR-box, one can
achieve the maximal possible violation of the CHSH inequality \cite{chsh}
(CHSH=4, compared to $2\sqrt{2}$ for quantum theory, or $\leq 2$ for
local classical theories).

We can also consider the anti-PR-box described by
\begin{equation} \
P_{\bar{PR}} (a b | x y) = \left\{
\begin{array}{ccl} \frac{1}{2} & : & a
\oplus b \neq x y\; \textrm{(mod 2)} \\
0& :  & \ot
\end{array} \right. ,
\end{equation}
Mixing PR-boxes ($s_0$) and anti-PR-boxes ($s_1$) we obtain a family
of noisy PR-boxes $s(p) \in S$ with probability distributions:
\begin{equation}
P_{s(p)}(a b |x y) = p P_{PR} (a b | x y) + (1-p) P_{\bar{PR}} (a b
| x y)
\end{equation}
Note that the `maximally mixed state' in which both outputs are
random and uncorrelated, is s(1/2).

There exists a twirling operation of all
devices (all probability distributions $P(a b | x y)$ \footnote{We would be happy to restrict to
non-signalling probability distributions, but in fact this is not
necessary, and we can prove the stronger result given.}) into $S$
\cite{lluis}.
To perform the twirling, the two parties generate three maximally
random shared bits $\alpha, \beta, \gamma$, and then perform local
transformations on the inputs and outputs of their terminals as
follows:
\begin{eqnarray}
x & \rightarrow & x \oplus \alpha \\
y & \rightarrow & y \oplus \beta \\
a & \rightarrow & a \oplus \beta x \oplus \alpha \beta \oplus \gamma \\
b & \rightarrow & b \oplus \alpha y \oplus \gamma
\end{eqnarray}
These transformations can be achieved either by re-labeling, or by
adding wires and gates to their local terminals. Note that these are all the local reversible
transformations that leave the PR-box rule $a \oplus b = xy$
invariant. It is straightforward to check that this achieves the
desired twirling.

We can consider any two-terminal device to be `separable' if the
probability distribution for its inputs and outputs can be
replicated by a local hidden variable model. I.e. if there exist
probability distributions $P_0(i)$, $P_A(a |x i)$, $P_B (b |y i)$
such that
\begin{equation} \label{sep_prob_dist}
P(a b | x y) = \sum_i P_{0}(i) P_A (a|x i) P_B (b |y i)
\end{equation}
We say a device is `entangled' if it is not separable. The family of
noisy-PR boxes defined above contains both entangled and separable
states, with an entanglement threshold of $p_s=3/4$.

We are now in a position to address the purification of noisy
PR-boxes. Given two noisy PR-boxes $(s(p)$ with $p>p_s)$, can we
produce a single PR-box with higher purity ($s(p')$ with $p'>p$)
using local operations (adding wires and gates to the local
terminals) and classical communication?

\begin{figure}
\epsfig{file=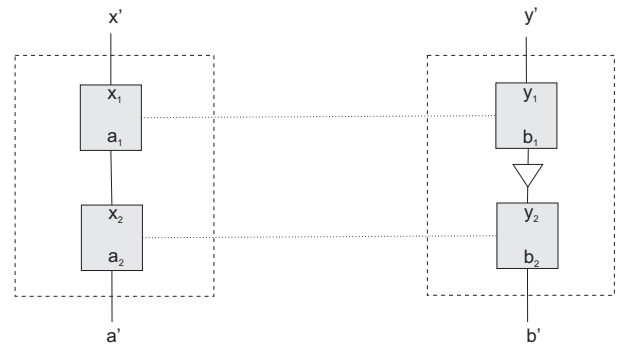, width=8cm} \caption{A possible (yet
unsuccessful) strategy for generating a PR-box of higher purity (the
dashed outer boxes) from two lower purity PR-boxes(the connected
shaded boxes), using local re-wirings (e.g. inputing $a_1$ into
$x_2$) and a NOT-gate.} \label{two-PR_fig}
\end{figure}

An example strategy is shown in figure $\ref{two-PR_fig}$, and it is
not obvious \emph{a priori} whether such a purification is possible.
Now, given the theory proved above, we know that it is not. Two
copies of a noisy PR-box cannot be purified by local wirings and
classical communication.

\smallskip
\textbf{Generalized probabilistic theories}: Both Werner states and
PR-boxes can be considered within a general probabilistic framework
\cite{barrett, hardy}, in which a state is characterised by the
joint probability distribution for some set of \emph{fiducial}
measurements on each subsystem, and the allowed states form a convex
set.

In the case of qubits in quantum theory, these fiducial measurements
can be taken to be measurements of the three Pauli operators
$\sigma_x, \sigma_y, \sigma_z $. The convex set of probability
distributions corresponding to allowed single qubit states will then
be isomorphic to the Bloch sphere. States comprised of multiple
qubits, such as the Werner states, can be completely characterised
by giving the joint outcome probabilities for every combination of
Pauli measurements on the subsystems.

For the PR-box, each terminal can be thought of as a primitive
subsystem with two binary-outcome fiducial measurements (represented
by the two possible inputs). The probability distribution $P_{PR}(a
b | x y)$ can then be understood as the joint probability of
obtaining outcomes $a$ and $b$ when fiducial measurements $x$ and
$y$ are performed on the subsystems. There are no quantum systems
that can be completely characterised by two binary outcome
measurements, and even if there were they could not generate the
non-locality inherent in the PR-box, so $P_{PR}(a b | x y)$ lies
outside the set of allowed quantum states. However, it can be
embedded within a different theory, called `generalised
non-signalling theory' (or `box-world'), in which all non-signalling
probability distributions for the fiducial measurements are allowed
states \footnote{A probability distribution is non-signalling if the
marginals for each party do not depend on the other parties'
measurement choices.}.

Once the set of allowed states has been fixed, all allowed
operations and non-fiducial measurements can be represented by
linear maps acting on the fiducial measurement probabilities
\cite{barrett}, and twirlability can be understood in terms of these
transformations. As in (\ref{sep_prob_dist}), a state is considered separable if it
can be represented by a convex combination of product states
(factorisable probability distributions in which both subsystems
have an allowed state), and entangled otherwise. It is easy to see that
separable states remain separable under LOCC.

All generalised probabilistic theories represented within such a framework satisfy
the three requirements given in the introduction.
Hence in all such theories, two copies of a noisy entangled
state cannot be deterministically purified if they come from a
twirlable family. Even if quantum mechanics were eventually to be
superseded by some different theory, so long as that theory could be
represented within the generalised probabilistic framework of
\cite{barrett}, this result would still hold.

For noisy PR-boxes (now interpreted as states in box-world), this
means that even if we could  perform operations on them that were
not wirings (e.g. a joint operation analogous to a local two-qubit
unitary), we would not be able to purify them.

\smallskip
\textbf{Conclusions:} We have shown that two copies of a noisy
entangled state, taken from some twirlable family, cannot be
transformed into a single copy of higher purity via LOCC. Due to the
very minimal requirements of the proof, this result will hold not just in quantum
theory, but in any theory that can be expressed within a generalised
probabilistic framework \cite{barrett}. In quantum theory, this
leads to the particular result that two copies of a Werner state
cannot be purified, and in box-world that two copies of a noisy
PR-box cannot be purified.

Interestingly, an analogous argument should apply to two-copy
purification of other properties (in place of entanglement), given a class of allowed
operations that cannot generate that property (instead of LOCC).

\acknowledgments The author would like to thank Anthony Laing, Sandu Popescu, Lluis Massanes, and Aram Harrow for helpful discussions.
This work was funded by a Royal Society University Research Fellowship, and was supported in part
by the EU's FP6-FET Integrated Projects
SCALA (CT-015714) and QAP (CT-015848).

\end{document}